\documentclass[]{ceurart}

\usepackage{amsmath,amssymb,amsfonts}
\usepackage{algorithmic}
\usepackage{graphicx}
\usepackage{textcomp}
\usepackage{multirow}
\usepackage{multicol}
\usepackage{hyperref}

\usepackage{color}
\usepackage{xspace}
\usepackage{adjustbox}
\usepackage[dvipsnames]{xcolor}
\usepackage{algorithmic}
\usepackage{colortbl}
\usepackage{tabu}
\usepackage{amsmath}
\usepackage{multibib}
\def\BibTeX{{\rm B\kern-.05em{\sc i\kern-.025em b}\kern-.08em
    T\kern-.1667em\lower.7ex\hbox{E}\kern-.125emX}}

\usepackage{subfig}

\usepackage{tikz}
\usepackage{amsmath}

\usepackage{multicol}
\usepackage{comment}
\usepackage{float}
\usepackage{csquotes}
\usepackage{textcomp}

\usepackage{filecontents}
\usepackage{xspace}
\usepackage{multirow}

\usepackage{multicol}
\usepackage{lipsum}
\usepackage{mwe}

\usepackage{framed}
\usepackage{paralist}
\definecolor{shadecolor}{RGB}{192,192,192}

\usepackage[ruled,linesnumbered]{algorithm2e}

\usepackage{listings}
\usepackage{float} 
\usepackage{placeins} 

\clearpairofpagestyles 
\ofoot{\thepage} 

\begin{document}
\copyrightyear{2024}
\copyrightclause{Copyright for this paper by its authors.
  Use permitted under Creative Commons License Attribution 4.0
  International (CC BY 4.0).}

\conference{SIGIR eCom'24: The 2024 SIGIR Workshop On eCommerce,
  July 18, 2024, Washington, D.C., USA}

\title{Next-Gen Sponsored Search: Crafting the Perfect Query with Inventory-Aware RAG (InvAwr-RAG) - Based GenAI}


\author[1]{Md Omar Faruk Rokon}[%
email=mdomarfaruk.rokon@walmart.com,
]
\cormark[1]
\address[1]{Walmart AdTech, Sunnyvale, CA, USA}

\author[1]{Weizhi Du}[%
email=weizhi.du@walmart.com,
]

\author[1]{Zhaodong Wang}[%
email=zhaodong.wang@walmart.com,
]

\author[1]{Musen Wen}[%
email=musen.wen@walmart.com,
]


  
\cortext[1]{Corresponding author.}

\begin{keywords}
  Dynamic Query Rewriting \sep
  Generative AI in Advertising \sep
  Sponsored Search \sep
  E-commerce Advertising \sep
  RAG \sep
\end{keywords}

\maketitle

\begin{abstract}
Sponsored search plays a crucial role in e-commerce revenue generation, where advertisers strategically bid on keywords to capture the attention of users through relevant search queries. However, the process of identifying pertinent keywords for a given query presents significant challenges because of a vast and evolving keyword landscape, ambiguous intentions, and topic diversity. This paper highlights an opportunity for to earn a considerable amount of Ads revenue and user engagement where a significant proportion of queries fail to retrieve any sponsored ads. To utilize this opportunity, we introduce the Inventory-Aware RAG-based Generative AI model (InvAwr-RAG), which integrates advanced semantic retrieval and real-time inventory data. This model combines dynamically generated and historically successful queries to align with available inventory and ad campaigns while diversifying rewritten queries to enhance relevance and user engagement. Preliminary results show a significant 68\% increase in fill rate and balanced relevance metrics, indicating a strong potential for increased ad revenue. The InvAwr-RAG model sets a new standard in dynamic query optimization, significantly improving ad relevancy, advertiser ROI, and user experience on Walmart's digital platform.
\end{abstract}


\vspace{-0.3cm}
\section{Introduction}
\vspace{-0.2cm}
\label{sec:intro}

Sponsored search is a cornerstone of revenue generation in e-commerce, where advertisers bid on keywords to display their ads in response to user queries. This system, however, faces significant challenges, including the alignment of user queries with relevant ads—a process complicated by the vast, dynamic keyword landscape and diverse user intents. In the competitive landscape of digital advertising, the efficiency of sponsored search systems is paramount for driving revenue and enhancing user experience on e-commerce platforms like Walmart. A significant challenge that Walmart faces is the presence of search queries that fail to retrieve any sponsored product ads—accounting for approximately 13\% of all searches. This issue represents a substantial revenue loss and a missed opportunity to engage potential customers. The inability to show relevant ads not only impacts Walmart’s bottom line but also diminishes the effectiveness of the platform for advertisers seeking visibility and for customers who may miss out on discovering products of interest. Hence, there is a compelling business need for a solution that can dynamically align search queries with available inventory and advertising goals, ensuring that every search can result in meaningful ad placements.

The core problem this research addresses is the high rate of search queries that yield no ad results due to mismatches between user queries and the current inventory or the specificities of real-time bidding budgets. The challenge is twofold: firstly, to enhance the relevance of ad placements to ensure they correspond with available inventory and meet advertiser bidding strategies; and secondly, to maintain or even improve user experience by presenting ads that are perceived as relevant and potentially interesting. This problem is crucial because it affects Walmart's ability to maximize ad revenue, utilize advertising space efficiently, and ensure customer satisfaction.

Current solutions in sponsored search fall into two main categories: information retrieval (IR) and generative or Natural Language Generation (NLG) based retrieval. IR methods, such as Dense Retrieval (DR) approaches including ANCE \cite{xiong2020approximate}, RocketQA \cite{qu2020rocketqa}, and NGAME \cite{dahiya2023ngame}, employ advanced deep learning models to create dense representations of queries and keywords, achieving state-of-the-art performance by utilizing effective negative mining strategies \cite{bai2018scalable, broder2008search, broder2007semantic, bhatia2016extreme}.
Conversely, NLG-based methods like CLOVER \cite{mohankumar2023unified, mohankumar2021diversity} and ProphetNet-Ads \cite{qi2020prophetnet} use generative models to transform user queries into more effective keywords, aiming to synthesize query forms that better match available ads.
Despite these advances, both IR and NLG methods often overlook real-time inventory data, leading to a disconnect between the generated queries and the ads available, resulting in many queries failing to retrieve any ads. This lack of synchronization with the dynamic inventory and ad campaign specifics renders these methods less effective and unable to leverage potential insights from immediate market conditions.


To bridge this gap, we introduce an innovative Inventory-Aware RAG-based Generative AI model (InvAwr-RAG) at Walmart. Our system leverages state-of-the-art technologies, including two-tower BERT embeddings for deep understanding of query and product semantics, combined with advanced indexing techniques for rapid and efficient retrieval of inventory data. By integrating large language models (LLMs) with Retrieval-Augmented Generation (RAG), our approach rewrites incoming user queries in real-time to align them with the most relevant and available sponsored products. Furthermore, our system enhances the query pool by blending dynamically generated queries with proven successful queries from historical search logs that consistently result in ad displays. This hybrid approach ensures that rewritten queries are immediately applicable and effective, reflecting live updates in inventory and adhering to the nuances of real-time ad bidding strategies.

This research introduces several significant innovations in the field of sponsored search:
\begin{itemize}
    \item Introducing a robust model for Dynamic Query Rewriting: Our model dynamically adjusts user queries to improve alignment with real-time inventory and advertiser bids, effectively turning previously unfulfillable searches into opportunities for ad placement.
    \item Enhancing Real-Time Data Integration: By integrating real-time inventory and bidding data, our system ensures that every search query can result in relevant and effective ad displays.
    \item Hybrid Query Generation: Combining AI-generated queries with historically successful queries allows for a rich mix of freshness and reliability in ad placements.
    \item Improving User Experience and Advertiser ROI: Our model not only enhances the user experience by providing relevant ad suggestions but also increases Return on Investment (ROI) for advertisers by maximizing the visibility of their products in relevant searches.
    \item Scalability and Adaptability: The use of advanced data handling techniques and the adaptability of LLMs to learn from vast amounts of data ensure that our solution is scalable and can continuously evolve with changing market dynamics.
\end{itemize}

Our model demonstrated a +68\% improvement in fill rate for queries that previously failed to retrieve ads, thereby reducing the number of no-result queries substantially.
These findings not only show a marked improvement in fill rates but also underscore the potential of an increase in ad-related revenue by up to \$1 billion over the next five years. and enhance the overall shopping experience on Walmart's e-commerce platform.

\section{Methodology}
\vspace{-0.2cm}
In this section, we describe the development and implementation of the Inventory-Aware RAG-based Generative AI model (InvAwr-RAG). Our objective is to dynamically rewrite user queries to enhance ad relevance and search efficiency on Walmart's e-commerce platform. The methodology encompasses several key stages: data preparation, model architecture design, training, and real-time query processing.

\subsection{Data Preparation}
Efficient data preparation is crucial for our model's success. It involves meticulous collection and annotation of data to capture the complex relationships between search queries, product titles, and user interactions.

\textbf{Data for Two-Tower BERT Model: }We compiled a dataset for the Two-Tower BERT model, pairing search queries with product titles and associated relevance scores, enhanced through human annotations to refine accuracy.

\textbf{Search Log Analysis for Popular Queries: }
Our search log analysis identified popular queries based on search frequency and ad impressions, guiding the RAG component to align ad content effectively with user preferences.

\textbf{Collection and Annotation of Rewritten Queries: }
Furthermore, we compiled search queries from our logs to identify potential alternative or revised queries. We selected queries that led to at least 500 clicks on the same items over a six-month period, as these reflect a high level of user engagement and intent. To ensure the relevance and uniqueness of these queries, we subjected them to a rigorous human evaluation process. We filtered out any queries that were deemed too similar to existing ones, retaining only those that provided distinct alternatives. The finalized set of revised queries was then utilized to train our Query Rewrite LLM, ensuring that the model learns from real-world, effective search patterns.


\subsection{Model Architecture and Training}
The InvAwr-RAG model combines supervised learning with advanced fine-tuning, tailored for the dynamic e-commerce environment.

\textbf{Supervised Learning with Two-Tower BERT}
The Two-Tower BERT model processes user queries and product titles separately, using embeddings to capture deep semantic meanings essential for effective matching.

\begin{figure}[ht]
\centering
\includegraphics[width=0.8\textwidth]{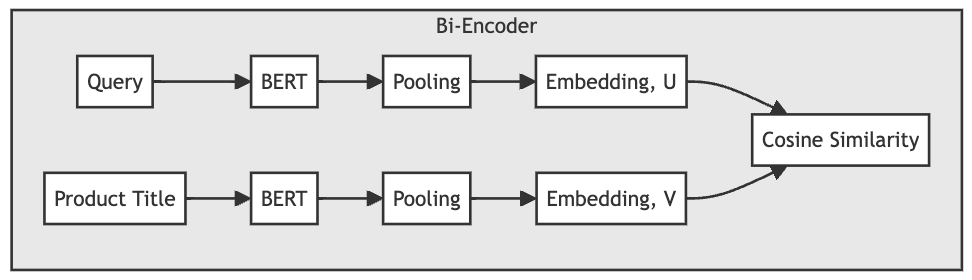}
\caption{The Two-Tower BERT model uses distinct encoders to generate embeddings for user queries (\( \text{Embedding, U} \)) and product titles (\( \text{Embedding, V} \)), enabling a nuanced match-making process. Through pooling operations, these embeddings are refined to represent the essential characteristics of the inputs. The system then uses cosine similarity to evaluate the degree of match between queries and products, thereby training the model to identify products most relevant to a given user query.}
\label{fig:two_tower_bert}
\end{figure}

\textbf{Fine-Tuning for Query Rewrite LLM with Low-Rank Adaptation (LoRA):}
Our Query Rewrite LLM is based on the Llama2 7B model, which we have fine-tuned using Low-Rank Adaptation (LoRA) to enhance its performance in the context of sponsored search query rewriting. The fine-tuning process involves optimizing the model with a specialized dataset to ensure it generates contextually relevant and inventory-aware rewritten queries. This approach allows us to leverage the extensive pre-training of Llama2 7B while adapting it specifically for our use case.  
The introduction of low-rank matrices \( A \) and \( B \) into the weight matrices optimizes the adaptation, minimizing the number of trainable parameters.

\begin{figure}[ht]
\centering
\includegraphics[width=0.8\textwidth]{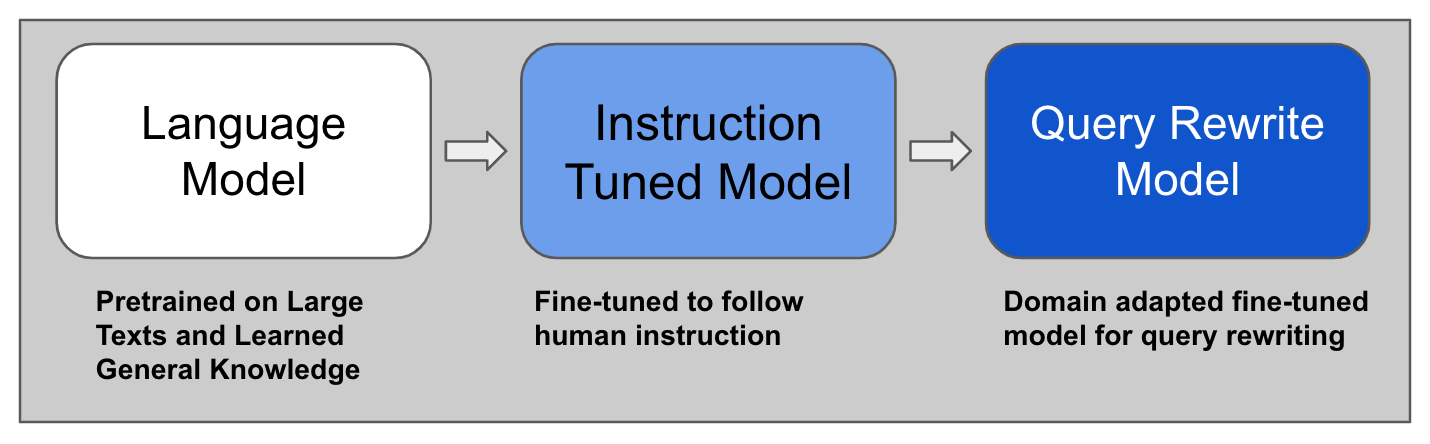}
\caption{The Query Rewrite LLM is enhanced through a two-pronged fine-tuning approach. Initially, the model learns to process and execute structured instructions. It is then further refined using Low-Rank Adaptation (LoRA), which allows for efficient, targeted adjustments to its architecture, enabling the generation of contextually relevant and inventory-aware query rewrites. This fine-tuning process ensures that the LLM can produce search queries that accurately reflect current inventory and user interests.}
\label{fig:llm_query_rewrite}
\end{figure}

This integrated approach ensures that our model adapts dynamically to user inputs and inventory data, providing an effective bridge between user expectations and available products.

\begin{figure}[ht]
\centering
\includegraphics[width=0.9\textwidth]{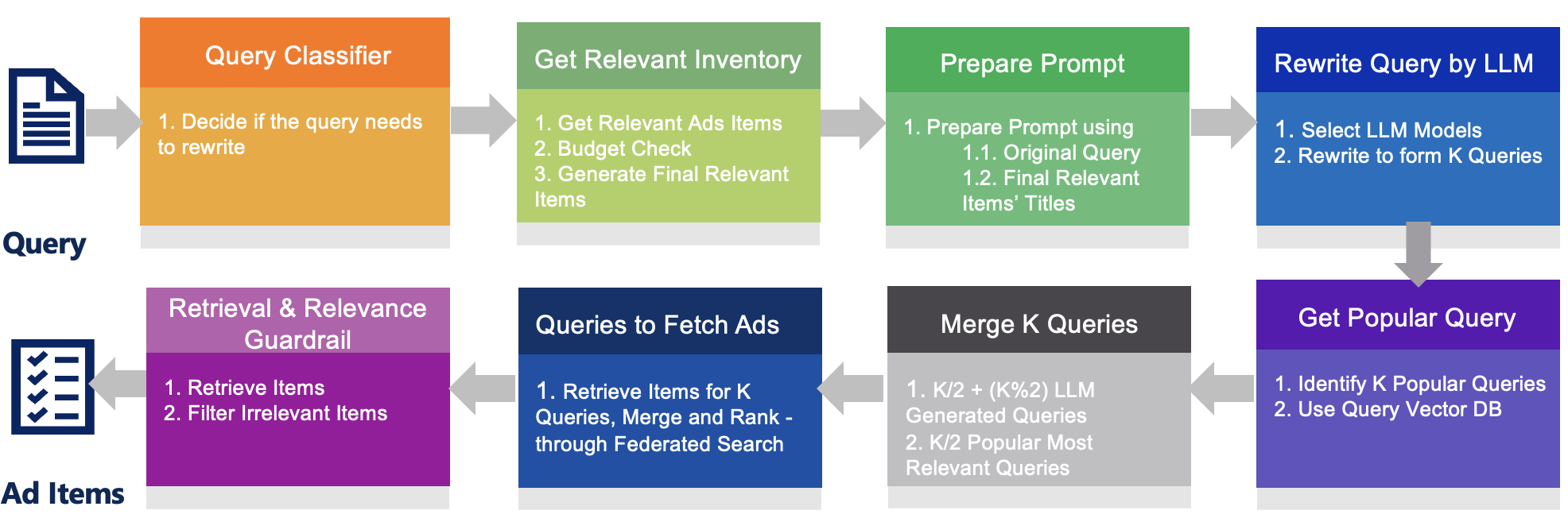}
\caption{Overview of our end-to-end RAG based query writing system.}
\label{fig:rag_steps}
\end{figure}

\subsection{Retrieval-Augmented Generation (RAG)}
Figure \ref{fig:rag_steps} provides a comprehensive overview of our end-to-end RAG-based query rewriting system. This system utilizes embeddings effectively to match user queries with relevant inventory items, considering both product relevance and budget constraints to ensure practical ad placements.

\textbf{Step1 - Query Classifier:} Our rule-based classifier assesses each user's search query to identify if it underperforms based on metrics like click-through rates. We redirect low-performing queries to a specialized route that enhances their retrieval effectiveness, optimizing our system's resource usage and improving outcomes for complex queries.

\textbf{Step2 - Dynamic Retrieval of Inventory Items:} We actively retrieve the top N inventory items from our vector database, selecting items based on their cosine similarity to the user's query while ensuring budget constraints are met.

\begin{figure}[ht]
\centering
\includegraphics[width=0.9\textwidth]{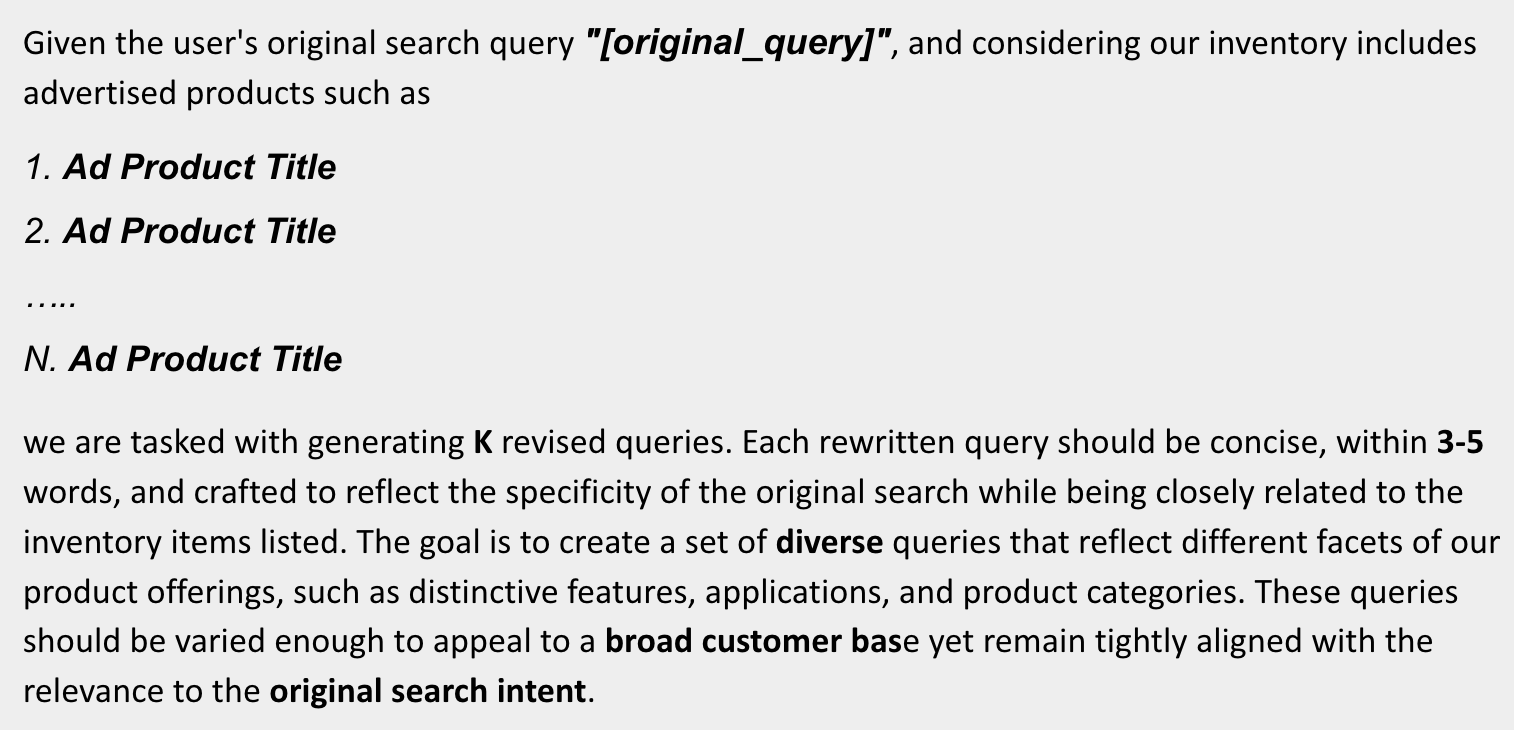}
\caption{Illustration of the LLM prompt setup for query rewriting. We combine the original user query with selected product attributes to form prompts that guide our generative model in producing diverse and relevant search queries.}
\label{fig:llm_query_prompt}
\end{figure}

\textbf{Step3 - Preparation of Query Rewrite Prompts:} After identifying relevant items, we prepare prompts using the original query combined with descriptions of these items. Our goal is to generate diverse yet relevant rewritten queries that cater to a broad range of customer interests without compromising relevancy, as shown in Figure \ref{fig:llm_query_prompt}.

\textbf{Step4 - Rewrite Query by LLM:} We use a generative model trained on query suggestion synthesis to produce rewritten queries that are not only relevant but also effectively linked to the available inventory.

\textbf{Step5 - Get Popular Queries:} We identify and integrate popular queries from our logs that resemble the rewritten queries generated by the LLM, enriching our query suggestions with common search terms and patterns.

\textbf{Step6 - Merge K Queries and Retrieval:} We merge the rewritten queries generated by the LLM with popular queries from our history logs. We then use these queries to dynamically retrieve relevant ad items. Our cross-encoder based BERT model evaluates the relevancy of each item, ensuring that only those meeting our predefined relevancy thresholds are displayed. This approach not only tailors ad displays to current search intents but also boosts customer satisfaction by aligning ads closely with user expectations.

Our comprehensive RAG system effectively bridges the gap between user queries and suitable products, optimizing ad displays and enhancing the shopping experience. By leveraging advanced algorithms for query processing and item retrieval, and ensuring that all interactions are guided by relevance and user intent, our system not only optimizes ad displays but also significantly enhances the user shopping experience. The careful integration of dynamic retrieval processes with a robust relevance framework ensures that our e-commerce platform can cater to a wide array of customer needs while maintaining high efficiency and scalability, ultimately leading to a more engaging and successful search experience for all users.

\section{Experiment and Results}
\vspace{-0.2cm}
\label{sec:experiment_results}

This section presents results for the InvAwr-RAG model, aimed at enhancing the effectiveness of Walmart's sponsored search system.

\subsection{Selection Criteria for N Items and K Queries}
An essential aspect of our methodology in the InvAwr-RAG model involves the selection of N=20 items and K=5 rewritten queries. These parameters were carefully determined based on both empirical evidence and operational efficiency, ensuring optimal performance and relevance.

\textbf{Determining N=20 Items: }
The choice of N=20 items for retrieval from our vector database is grounded in the following considerations:

{\em 1. Diversity and Coverage: } Retrieving 20 items allows the system to achieve a balance between diversity and specificity. This number is large enough to cover various aspects of user queries, yet manageable enough to maintain high relevance and avoid overwhelming the user with options.
    
{\em 2. User Experience}: Based on user interaction data, we observed that presenting up to 20 items maximizes engagement without causing decision fatigue. Users are more likely to browse through and interact with a set of 20 well-curated product suggestions.
    
{\em Computational Efficiency: } From a technical perspective, retrieving 20 items strikes an optimal balance between computational load and response time, ensuring that the system remains responsive even under high traffic conditions.

\textbf{Choosing K=5 Rewritten Queries: }
The decision to generate K=5 rewritten queries for each original query was based on several factors:

{\em 1. Query Variation}: Five rewritten queries provide sufficient variation to explore different linguistic formulations and product matches, increasing the chances of hitting upon the most effective phrasing that captures the user's intent.
    
{\em 2. Precision and Focus}: Limiting the number of rewritten queries to five helps maintain focus and precision in query suggestions, ensuring each query is highly targeted and likely to yield relevant results.


\subsection{Offline Results}
We conducted an initial offline evaluation on a set of 10,000 queries historically known to have a 0\% fill rate—queries that consistently failed to retrieve any ad items. Our fine-tuned Llama2 7B model, integrated into the InvAwr-RAG system, achieved a 68\% fill rate on this challenging set, demonstrating substantial improvement in query coverage and relevance.

For comparison, we also evaluated the performance of queries rewritten by GPT-4, a state-of-the-art LLM, under similar conditions. GPT-4 achieved a fill rate of 53\%, which, while impressive, underscores the additional benefits gained from our fine-tuning approach that specifically addresses the retail context of Walmart.

\begin{table}[ht]
\centering
\begin{tabular}{lcc}
\hline
\textbf{Model} & \textbf{Fill Rate} & \textbf{NDCG@8} \\
\hline
Baseline (0\% Fill Rate) & 0\% & 0 \\
GPT-4 & 53\% & 0.6458 \\
InvAwr-RAG & 68\% & 0.6847 \\
\hline
\end{tabular}
\caption{Preliminary comparison of fill rates and NDCG scores for a set of historically zero-return queries.}
\label{table:fill_rates}
\end{table}

To further assess the relevancy of the returned ad items to the original queries, we utilized the NDCG metric at a cutoff of 8 (NDCG@8). This metric evaluates the quality of the ranking by measuring the usefulness, or gain, of the ad items based on their positions in the result list. Higher scores indicate better relevancy. The evaluations were conducted by third-party human evaluators who assessed the top 8 items returned by each model. As shown in the table, our InvAwr-RAG model not only improved fill rates but also demonstrated superior relevance as indicated by its higher NDCG score compared to both the baseline and GPT-4.

These preliminary findings suggest that the InvAwr-RAG model has the potential to significantly enhance the sponsored search system, making a strong case for further investigation through planned A/B testing.

We will conduct A/B testing to measure the InvAwr-RAG model’s impact in a live setting, focusing on Fill Rate, Click-Through Rate, Conversion Rate, and Revenue Impact. These metrics will help validate the model's potential to transform ineffective queries into profitable engagement opportunities, with preliminary data already indicating significant improvements.




\section{Discussion}
\vspace{-0.2cm}
\label{sec:discussion}

This section clarifies the essential role of query rewriting in enhancing item retrieval from the vector database and addresses potential queries about the confidence in rewritten queries and the integration of popular searches.

\textbf{The Complementary Role of Query Rewriting: }
Query rewriting is vital even with a capable vector database because about 13\% of queries yield no results. It bridges the gap by transforming queries into formats more likely to match the inventory, improving relevance and user intent alignment. Rewriting also adjusts to dynamic inventory changes and diverse user expressions, ensuring robust and responsive search capabilities.

\textbf{Confidence in Rewritten Queries: }
Confidence in our rewritten queries is founded on a robust combination of contextual knowledge, historical data, and human annotation. Our LLM's training, enriched with extensive contextual and industry-specific data, enables the generation of pertinent queries even when initial retrievals are unsuccessful. Historical data anchors the model's suggestions in proven past interactions, while human annotation refines our dataset with alternative query expressions that have historically led to successful outcomes. This methodical approach ensures the efficacy of the model across various scenarios.

\textbf{Incorporating Popular Queries}
Integrating popular queries plays a crucial role by reflecting collective user behavior, which ensures that rewritten queries resonate with broad user search patterns. This strategy not only captures current trends, providing timely relevance, but also combines the real-time adaptability of LLM-generated queries with the solid foundation of user preferences. This hybrid approach is particularly valuable during peak shopping periods and market shifts, effectively guiding the LLM to produce queries that align with the latest user intents and market trends. This strategic integration enhances the relevance and effectiveness of our search system, benefiting both users and the platform.

\section{Conclusions}
\vspace{-0.2cm}
\label{sec:concl}

This research has demonstrated the effectiveness of the Inventory-Aware RAG-based Generative AI model (InvAwr-RAG) in addressing significant inefficiencies in sponsored search systems on e-commerce platforms like Walmart. By dynamically rewriting queries to align with real-time inventory and ad campaigns, the InvAwr-RAG model significantly reduces the occurrence of no-result queries and has the potential to increase ad-related revenue substantially.

Preliminary results of our system have shown promising results. These findings highlight the potential of integrating advanced AI technologies to enhance the relevance and effectiveness of ad placements, thereby improving both user experience and advertiser ROI.

The forthcoming A/B test will provide a more definitive analysis of the InvAwr-RAG model's performance. Beyond this, future work will focus on refining the model’s understanding of user intent and expanding its applicability across more diverse product categories and bidding strategies. Continual improvements in scalability and efficiency will also be critical as Walmart’s inventory and user base expand. 


\bibliography{BIB/rokon}

@inproceedings{bai2018scalable,
  title={Scalable query n-gram embedding for improving matching and relevance in sponsored search},
  author={Bai, Xiao and Ordentlich, Erik and Zhang, Yuanyuan and Feng, Andy and Ratnaparkhi, Adwait and Somvanshi, Reena and Tjahjadi, Aldi},
  booktitle={Proceedings of the 24th ACM SIGKDD international conference on knowledge discovery \& data mining},
  pages={52--61},
  year={2018}
}

@inproceedings{broder2008search,
  title={Search advertising using web relevance feedback},
  author={Broder, Andrei Z and Ciccolo, Peter and Fontoura, Marcus and Gabrilovich, Evgeniy and Josifovski, Vanja and Riedel, Lance},
  booktitle={Proceedings of the 17th ACM conference on information and knowledge management},
  pages={1013--1022},
  year={2008}
}

@inproceedings{broder2007semantic,
  title={A semantic approach to contextual advertising},
  author={Broder, Andrei and Fontoura, Marcus and Josifovski, Vanja and Riedel, Lance},
  booktitle={Proceedings of the 30th annual international ACM SIGIR conference on Research and development in information retrieval},
  pages={559--566},
  year={2007}
}

@article{xiong2020approximate,
  title={Approximate nearest neighbor negative contrastive learning for dense text retrieval},
  author={Xiong, Lee and Xiong, Chenyan and Li, Ye and Tang, Kwok-Fung and Liu, Jialin and Bennett, Paul and Ahmed, Junaid and Overwijk, Arnold},
  journal={arXiv preprint arXiv:2007.00808},
  year={2020}
}

@article{qu2020rocketqa,
  title={RocketQA: An optimized training approach to dense passage retrieval for open-domain question answering},
  author={Qu, Yingqi and Ding, Yuchen and Liu, Jing and Liu, Kai and Ren, Ruiyang and Zhao, Wayne Xin and Dong, Daxiang and Wu, Hua and Wang, Haifeng},
  journal={arXiv preprint arXiv:2010.08191},
  year={2020}
}

@inproceedings{dahiya2023ngame,
  title={Ngame: Negative mining-aware mini-batching for extreme classification},
  author={Dahiya, Kunal and Gupta, Nilesh and Saini, Deepak and Soni, Akshay and Wang, Yajun and Dave, Kushal and Jiao, Jian and K, Gururaj and Dey, Prasenjit and Singh, Amit and others},
  booktitle={Proceedings of the Sixteenth ACM International Conference on Web Search and Data Mining},
  pages={258--266},
  year={2023}
}

@article{bhatia2016extreme,
  title={The extreme classification repository: Multi-label datasets and code},
  author={Bhatia, Kush and Dahiya, Kunal and Jain, Himanshu and Mittal, Anshul and Prabhu, Yashoteja and Varma, Manik},
  journal={URL http://manikvarma. org/downloads/XC/XMLRepository. html},
  year={2016}
}

@inproceedings{mohankumar2021diversity,
  title={Diversity driven query rewriting in search advertising},
  author={Mohankumar, Akash Kumar and Begwani, Nikit and Singh, Amit},
  booktitle={Proceedings of the 27th ACM SIGKDD Conference on Knowledge Discovery \& Data Mining},
  pages={3423--3431},
  year={2021}
}

@inproceedings{mohankumar2023unified,
  title={Unified Generative \& Dense Retrieval for Query Rewriting in Sponsored Search},
  author={Mohankumar, Akash Kumar and Dodla, Bhargav and K, Gururaj and Singh, Amit},
  booktitle={Proceedings of the 32nd ACM International Conference on Information and Knowledge Management},
  pages={4745--4751},
  year={2023}
}

@inproceedings{qi2020prophetnet,
  title={Prophetnet-ads: A looking ahead strategy for generative retrieval models in sponsored search engine},
  author={Qi, Weizhen and Gong, Yeyun and Yan, Yu and Jiao, Jian and Shao, Bo and Zhang, Ruofei and Li, Houqiang and Duan, Nan and Zhou, Ming},
  booktitle={Natural Language Processing and Chinese Computing: 9th CCF International Conference, NLPCC 2020, Zhengzhou, China, October 14--18, 2020, Proceedings, Part II 9},
  pages={305--317},
  year={2020},
  organization={Springer}
}

\end{document}